# Molecular Arrangements in the First Monolayer of Cu-Phthalocyanine on In$_2$O$_3$(111)

Matthias Blatnik[1,2], Fabio Calcinelli[3], Andreas Jeindl[3], Moritz Eder[1], Michael Schmid[1], Jan Čechal[2], Ulrike Diebold[1], Peter Jacobson[1,4], Oliver T. Hofmann[3] and Margareta Wagner[1*]

[1] *Institute of Applied Physics, TU Wien, 1040 Vienna, Austria*
[2] *CEITEC – Central European Institute of Technology, Brno University of Technology, 612 00 Brno, Czech Republic*
[3] *Institute of Solid State Physics, Graz University of Technology, 8010 Graz, Austria*
[4] *School of Mathematics and Physics, University of Queensland, 4068 St. Lucia, Australia*

[*] Corresponding author: margareta.wagner@tuwien.ac.at

**Abstract**

Well-ordered organic molecular layers on oxide surfaces are key for organic electronics. Using a combination of scanning tunneling microscopy (STM) and non-contact atomic force microscopy (nc-AFM) we probe the structures of copper phthalocyanine (CuPc) on In$_2$O$_3$, a model for a prototypical transparent conductive oxide (TCO). These scanning-probe images allow the direct determination of the adsorption site and distortions of the molecules, which are corroborated by DFT calculations. Isolated CuPc molecules adsorb in a flat, slightly tilted geometry in three symmetry-equivalent configurations on the stoichiometric In$_2$O$_3$(111) surface. Increasing the coverage leads to densely-packed 1D chains oriented along $\langle 1\bar{1}0 \rangle$ directions, which dissolve into a highly ordered (2 × 2) superstructure upon increasing the CuPc density to ¾ per surface unit cell. At a coverage of one CuPc per surface unit cell, a densely packed (1 × 1) superstructure fully covers the surface. The molecules still assume the same site and orientation as before, but they partially overlap to accommodate the high packing density, leading to a bending of the molecules. These results are compared to the behavior of CoPc on In$_2$O$_3$(111). In summary, we demonstrate that a uniform first layer of metal-phthalocyanine molecules can be realized on the In$_2$O$_3$(111) surface when using the proper metal atom in the molecule.

# Introduction

Well-ordered layers of organic molecules on surfaces play a crucial role in advancing device technology, particularly in organic electronics and optoelectronics. While PEDOT:PSS (Poly(3,4-ethylenedioxythiophene) polystyrene sulfonate) has achieved widespread success as a polymer-based injection layer, its stability issues, notably acidity, water uptake, heat and UV light, can degrade devices over time.[1–3] Small molecules are often an attractive alternative to polymers as they are available as higher purity source materials. They are suitable for thermal evaporation as well as deposition from solution, and can form well-defined interfaces with tailored electronic structures. However, often only scarce knowledge is available on the surface ordering of these small molecules on technologically relevant transparent electrode materials.

A particularly interesting substrate is the strongly *n*-doped indium tin oxide (ITO, tin-doped indium oxide), a technologically relevant transparent conductive oxide (TCO) and one of the most utilized wide-band gap TCOs. This material finds extensive use as an anode in optoelectronic applications and electrochemistry, including ITO-phthalocyanine interfaces.[4–7] Its high transparency in the visible range of light and low electrical resistivity surpasses AZO (aluminum-doped zinc oxide) or GZO (gallium-doped zinc oxide),[8] and ITO is therefore commonly found as anode material in (organic) solar cells, OLEDs, and LCD screens, as well as in smart windows.[9,10] The bulk lattice constant of ITO is slightly compressed with respect to pure $In_2O_3$, but both materials share the same lattice symmetry (bcc, bixbyite) and their (111) surface can be described as the relaxed, bulk-terminated structure of $In_2O_3$.[11,12] In its pure form, $In_2O_3$ has a fundamental band gap of ≈2.9 eV,[13] and is intrinsically *n*-doped[14] with the conduction band minimum lying close to the Fermi level. With additional doping, the conduction band can shift below the Fermi level, leading to an electron accumulation layer close to the surface.[14–16]

On metal electrode surfaces, the organic semiconductor usually self-assembles into well-ordered structures. Key properties such as the metal-molecule interaction,[17] charge transfer[18] and energy level alignment[19,20] are relatively well understood. In contrast, molecular growth on wide-band gap semiconducting oxide surfaces is far less explored.[21,22] Oxide surfaces, in general, feature undercoordinated metal/oxygen atoms with specific reactivity towards molecular functional groups.[23,24] The larger unit cells of oxide surfaces compared to metals and a more corrugated topography and complex potential landscape can be beneficial for hosting larger organic molecules. However, it can also constrain adsorption geometries, which hinders or limits the surface diffusion necessary for self-assembly.[25] Therefore, selecting a molecule with appropriate functional groups, size, and electronic properties, is crucial for achieving an optimal self-assembly process on a specific substrate surface. Here, we focus on the molecular self-assembly and formation of the first monolayer of copper phthalocyanine (CuPc) on $In_2O_3$, a prototypical charge injection system.

The organic semiconductor molecule CuPc (Cu-$(C_8H_4N_2)_4$) is a 2-dimensional metal-organic compound consisting of four isoindole units (benzo-fused pyrrole) linked by nitrogen atoms (see structure superimposed on the AFM image in Figure 1b). The central cavity of the molecule is framed by the four pyrroles, capable of accommodating a $Cu^{2+}$ ion. The $Cu^{2+}$ replaces two hydrogen atoms found in the metal-free phthalocyanine base (2HPc), resulting in vibrantly blue-colored CuPc, which finds use as an organic semiconductor in electronic and photovoltaic devices, catalysis,[26,27] gas sensing, and, potentially, as an S=½ system in quantum applications.[28,29] Owing to their technological relevance, phthalocyanines have been investigated in a bottom-up approach from adsorption of individual molecules to island and layer formation. Current research on surface structures formed by phthalocyanines predominantly concentrates on $TiO_2$[30–38] and ZnO,[39–42] while fewer studies address various other oxides, including FeO,[43,44] $Al_2O_3$,[45–47] CoO,[48,49] $V_2O_3$,[50] STO,[49] $In_2O_3$,[25,52] and a few other materials.[53,54] The formation of a well-ordered layer is, however, scarce, and in the case of $In_2O_3$(111) depends on the metal ion in the center of the phthalocyanine. This will be discussed here for CuPc in comparison to CoPc.[25]

In this work we report on the growth of CuPc on single-crystalline $In_2O_3$(111) up to the formation of the first, densely-packed monolayer (ML). A combination of scanning tunneling microscopy (STM), non-contact atomic force microscopy (nc-AFM), and density functional theory (DFT) calculations reveals that CuPc lies flat on the stoichiometric $In_2O_3$(111) surface and forms highly ordered and large-area domains. While the molecules avoid overlap at intermediate coverages below 1 ML by first forming chains and later a (2 × 2) superstructure at a coverage of ¾ ML, we find overlap between adjacent isoindole arms in the (1 × 1) structure at a coverage of one CuPc per substrate unit cell. The molecule assumes the same adsorption site from low coverages, where it is an isolated entity, up to the densely packed (1 × 1) structure, where it adjusts to its surroundings by bending.

**Experimental and computational methods**

The experiments were conducted in a two-chamber UHV system (Omicron) consisting of the analysis chamber (base pressure $5 \times 10^{-11}$ mbar) and a preparation chamber ($1 \times 10^{-10}$ mbar). The STM and nc-AFM measurements were performed with a commercial Omicron low-temperature STM/AFM cooled to 4.7 K or 77 K using qPlus sensors[55] with a separate wire for the tunneling current and a differential preamplifier. Electrochemically etched W tips (20 μm wire diameter before etching) were glued to the tuning fork and initially cleaned in UHV by self-sputtering and field emission.[56] The images were acquired with three different sensors: (1) $k \approx 5{,}400$ N/m, $f_R \approx 77$ kHz, $Q \approx 74{,}000$, (2) $k \approx 2{,}000$ N/m, $f_R \approx 29$ kHz, $Q \approx 17{,}000$, (3) $k \approx 2{,}000$ N/m, $f_R \approx 22.8$ kHz, $Q \approx 4{,}900$. For the STM and AFM experiments, the tip was prepared on CuPc/In$_2$O$_3$(111) by gentle voltage pulses until a frequency shift less than −4 Hz was observed. The tip termination in AFM is most likely a fragment of a CuPc molecule resulting in a contrast similar to a CO-functionalized tip.[57,58] The data presented here are constant-current STM images and constant-height non-contact AFM measurements. The imaging contrast in AFM in the regime of short-range forces is characterized by attractive (dark, strongly negative frequency shift) and repulsive (bright, less negative or positive frequency shift) interactions with the AFM tip.

Indium oxide single crystals,[59] cut and polished along the (111) plane, are used as substrates. A photograph of the sample used in this work is shown in the Supplementary Information (Fig. S1). The samples are nominally undoped and exhibit n-type electronic properties in scanning tunnelling spectroscopy (STS) with a band gap of ≈3 eV and the Fermi level located at the bottom of the conduction band as expected for this material (see scanning tunneling spectroscopy spectrum in Fig. S3 of the Supplementary Information). The crystal surface was prepared by several cycles of sputtering and annealing[12]. Sputtering was carried out in normal incidence with an ion gun (SPECS) scanning the crystal surface (1 keV Ar$^+$, ≈1.6 μA/cm$^2$ sample current, 5–10 min). Post-annealing at ≈450 °C took place under oxidizing conditions by backfilling the chamber with ≈6 × 10$^{-7}$ mbar O$_2$; this pressure was kept until the crystal was cooled to ≈150 °C. The cleanliness of the In$_2$O$_3$(111) was investigated with STM and LEED. A LEED pattern of the pristine surface as well as Fourier transforms of the STM images and STS measurements, are included in the Supplementary Information (Fig. S1 and Fig. S3). These data confirm the crystallinity and long-range ordering of the surface. Copper(II) phthalocyanine (Sigma-Aldrich, sublimed grade) was deposited by thermal sublimation at ≈400 °C (measured at the opening of the crucible) using a water-cooled Omnivac four-pocket evaporator, while the sample was kept at room temperature. After the deposition, the sample was post annealed at 200 °C to desorb any co-adsorbed water and enhance the diffusion of the molecules. Alternatively, keeping the sample at 200 °C during deposition without post-annealing yielded the same results.

Computational results were obtained with the FHI-aims code[60] using the provided default tight basis set. The In$_2$O$_3$(111) periodic slab was simulated using four O$_{12}$-In$_{16}$-O$_{12}$ trilayers, with a total thickness of 1.1 nm. The adsorbate was placed on one side of the slab only, and the repeated-slab-approach was employed with 3.8 nm vacuum and a dipole correction[61] to electrostatically decoupled periodic replicas on the z-axis perpendicular to the surface. Geometries were pre-relaxed using the GGA-level PBE-exchange correlation functional[62] and subsequently re-optimized using the HSE06 hybrid functional.[63] In both cases, a non-local many-body dispersion correction[64] was employed. All geometries were optimized by constraining the atoms of the bottom two $In_2O_3$ trilayers and relaxing the atoms of the molecules and the topmost trilayer until the maximum remaining force dropped below 0.01 eV/Å for all atoms in the unit cell. The geometry optimizations were conducted on a Γ-centered k-grid with $2 \times 2 \times 1$ k-points. All calculations were executed spin-unrestricted with a broadening of 0.03 eV and a self-consistent field (SCF) convergence threshold for the total energy of $1 \times 10^{-5}$ eV. Where relevant, doping of the substrate was simulated using the virtual crystal approximation.[65,66] The relaxed structures of individual CuPc molecules on $In_2O_3$(111) and the (1 × 1) structure formed at a coverage of 1 ML can be found in the Supplementary Information.

All computational results are reported at the HSE06 level unless explicitly mentioned otherwise. AFM simulations were performed using the ppafm package,[67] which employs a probe-particle model.[68] Here, the probe interacts with the DFT-obtained geometry using a Lennard-Jones potential and with the DFT-obtained electrostatic potential assuming an s-type tip and a charge of –0.1 e. The full set of parameters used for the simulation is reported in the Supplementary Information. AFM images were calculated over a range of tip heights and those showing the best visual agreement was chosen.

## Results

### Isolated CuPc molecules

Figure 1 gives an overview of the different structures observed for CuPc/$In_2O_3$(111) as function of coverage, displaying empty-states STM (top row) and nc-AFM images (bottom row). Isolated single molecules on terraces are observed at low coverages. With increasing coverage, they first assemble into short molecular chains, and, at coverages of 0.75 ML and 1 ML, into a (2 × 2) and a (1 × 1) superstructure, respectively. We thereby define 1 ML as 1 CuPc molecule per substrate unit cell, i.e., a density of $5.64 \times 10^{13}$ CuPc cm$^{-2}$. In the following, the individual structures introduced in Figure 1 are discussed. Overview STM images for different CuPc coverages and their Fourier transform are in the Supplementary Information (Fig. S6).

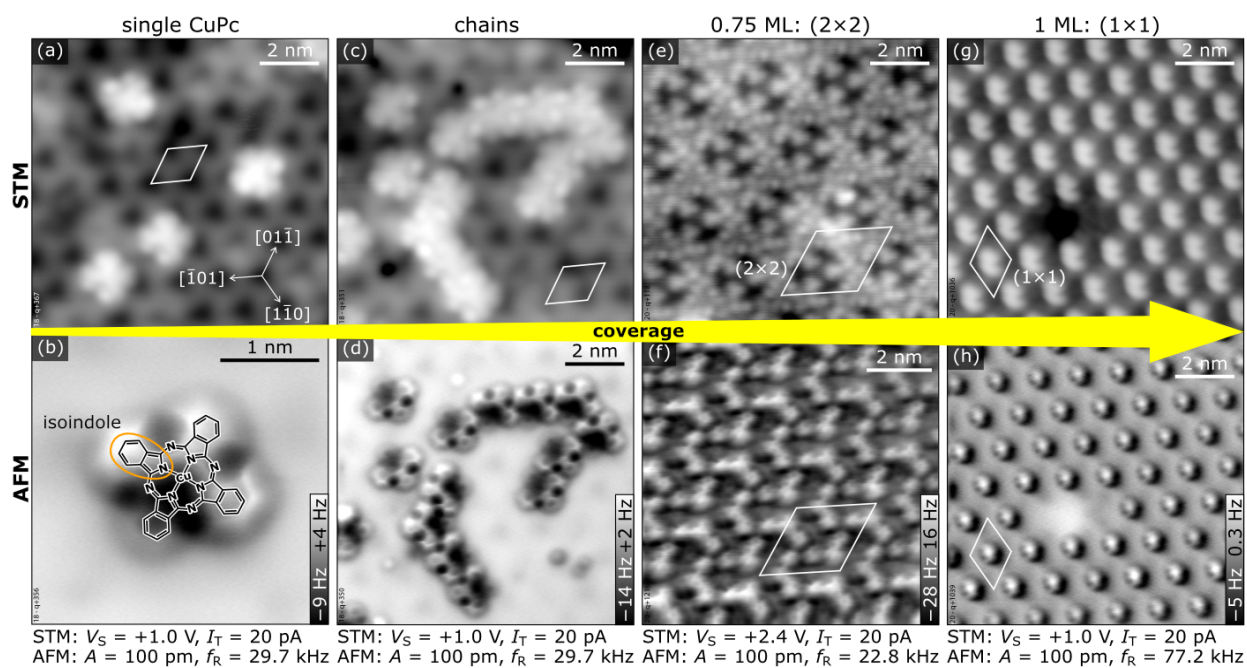

Fig. 1: CuPc structures on In$_2$O$_3$(111) as function of coverage imaged with STM (upper row) and AFM (bottom row). Starting with individual molecules (a, b), CuPc arranges into molecular chains (c, d), followed by a (2 × 2) superstructure formed by 0.75 ML (e, f), and the (1 × 1) structure at 1 ML of CuPc (g, h). The acquisition parameters for both STM and AFM images are provided at the bottom for each coverage. (a-d, g and h) were acquired at 4K, (e, f) at 80 K.

The adsorption of isolated CuPc molecules is displayed in Figure 2. The STM (Figure 2a) and AFM images (Figure 2b) display the same surface area. In STM, the single CuPc molecules appear as flower-like objects with 8 bright lobes at the periphery, two for each benzene ring of the molecule. This reflects the shape of the LUMO, with nodal planes bisecting the benzene rings.[69] The molecules adsorb in a seemingly flat-lying geometry (with an apparent diameter of ≈1.7 nm comparable to the gas-phase diameter), randomly distributed on the terraces of the substrate and without preference for defects or step edges (see Fig. S7 in the Supplementary Information). In accordance with the p3 symmetry of the substrate, three rotational orientations of the molecules are observed, where one long axis of the molecule is aligned in a particular ⟨1$\bar{1}$0⟩ direction. In-between the molecules, the In$_2$O$_3$(111) surface shows a periodic pattern of dark features in empty-states STM, where the surface is terminated by In(6c) and O(3c) atoms (yellow triangle in Figure 2a and 2d).[12]

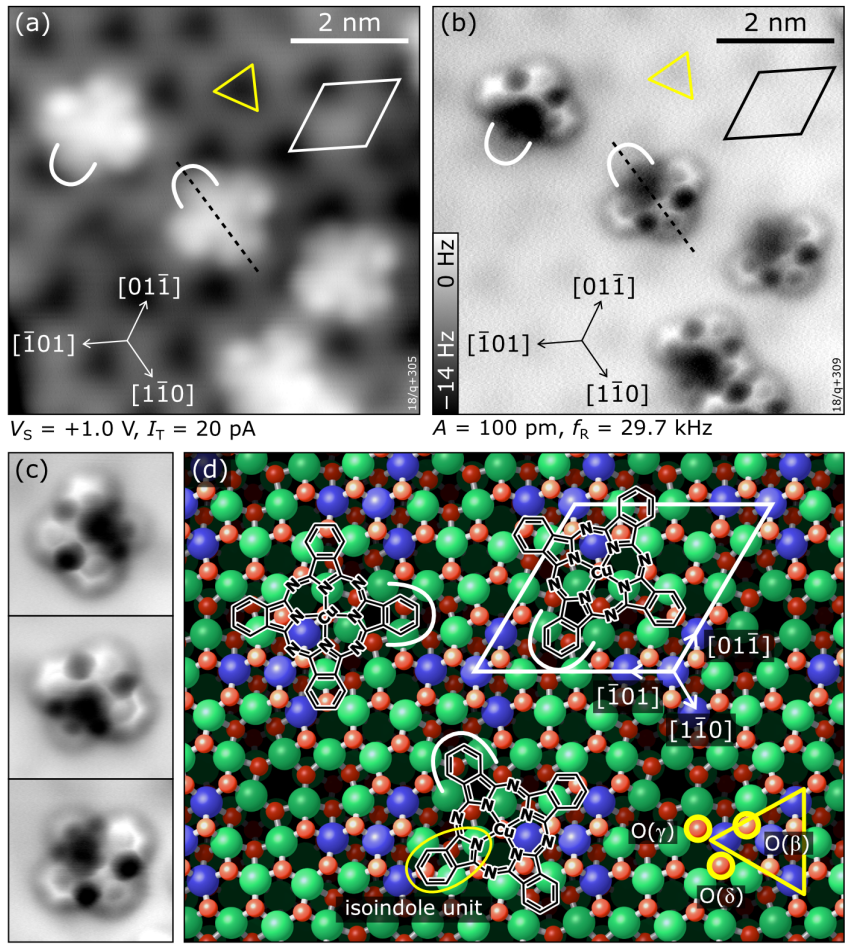

Fig. 2: Isolated CuPc molecules adsorbed on $In_2O_3(111)$. (a) STM image showing CuPc molecules in different rotational orientations. The substrate unit cell is indicated in white. The "dark" region of the unit cell (located at the corners of the unit cell) comprises In(6c) surface atoms and is highlighted by a yellow triangle. (b) Constant-height AFM image taken at the same spot. (c) Magnification of individual molecules with different orientations measured in constant-height AFM; each image is 2.5 × 2.5 $nm^2$ in size. (d) Adsorption site of CuPc on $In_2O_3(111)$: molecular structure of CuPc superimposed on the atomic model of the bare $In_2O_3(111)$ surface as observed in (a) and (b); the isoindole unit tilted towards the surface, which appears darker in AFM, is marked by a white curve. The adsorption site is described in the text. The surface unit cell is highlighted in white. The yellow triangle marks the arrangement of In(6c) atoms (blue); In(5c) are green, O(4c) are dark red. The three inequivalent O(3c) atoms (light red) with bonds to the In(6c) are labelled β, δ, and γ; note that three of each type are symmetrically arranged around the yellow triangle due to the 3-fold symmetry of the surface. All images were acquired at 5 K.

While STM is sensitive mostly to electronic states of the molecules, nc-AFM reveals geometric properties of the adsorbate.[57] In the constant-height AFM image of Figure 2b, three benzene rings per molecule are clearly discerned, together with the –N= bridge linking the isoindole groups (see Figure 1b). The fourth benzene fades into the substrate and must be, therefore, located significantly closer to the surface than the rest of the molecule, implying a slightly bent geometry of the whole molecule. The center of the molecule with the Cu atom is attractive to the tip (dark contrast). Due to their pronounced asymmetric appearance in AFM, the three different rotational orientations of the molecule are easily discerned; high-magnification images of these three orientations are shown in Figure 2c. In these detailed views also the fourth benzene ring is clearly observable, and partially also the internal structure related to the 5-membered rings of the isoindole units. Combining the information obtained from STM and nc-AFM measurements, it is possible to determine the adsorption site of the molecule as well as the rotational orientation of the CuPc molecule with respect to the surface unit cell (see Fig. S4 and Fig. S5 in the Supplementary Information). Figure 2d shows a sketch of the bulk-terminated, relaxed $In_2O_3(111)$ surface with the structure of the molecule superimposed at the position derived from the STM/AFM results. To rationalize the adsorption site, a brief introduction to the atomic structure of $In_2O_3(111)$ is necessary: The $In_2O_3(111)$ surface can be understood as an O(3c)-In(5c,6c)-O(4c) trilayer. Its four 6-fold coordinated In(6c) atoms per unit cell arrange in a three-pointed star (blue spheres highlighted by a yellow triangle in Figure 2d). The central In(6c) is connected to the three surrounding In(6c) via 3-fold coordinated O(β) atoms, and each of the surrounding In(6c) is connected via an O(γ) atom (just outside the corners of the yellow triangle in Figure 2d) to 5-fold coordinated In(5c) atoms. The third O atom bound to the surrounding In(6c) is called O(δ) in our nomenclature.[70] Returning to CuPc, a detailed analysis of the nc-AFM images shows that the central Cu ion of the molecule is on top of an O(γ). Further, the molecule is oriented such that one of the N atoms binding to the Cu points to the O(δ) atom next to the O(γ). Thus, two opposite isoindole units (including the one bent towards the surface, marked with a white curve in Figure 2d) are aligned with a $\langle 1\bar{1}0 \rangle$ direction of the substrate, and the benzene rings of the remaining two isoindole units are roughly located above O(δ) atoms at the periphery of the molecule. The isoindole unit bent towards the surface is located above an area without topmost O(3c) atoms (see Figure 2d), while the rest of the molecule is situated on O(3c) atoms. Thus, the curvature of the molecule follows the topography of the unit cell.

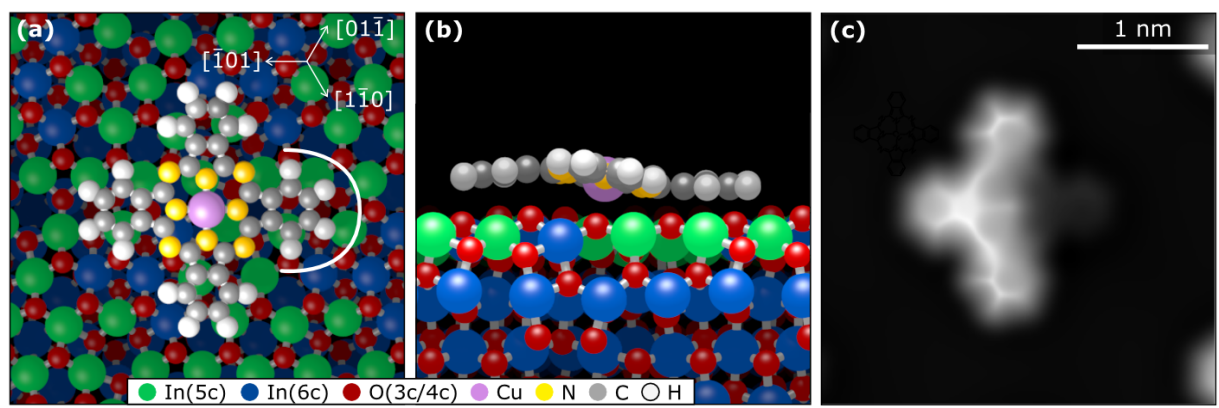

Figure 3. A single CuPc molecule on $In_2O_3(111)$. Top (a) and side (b) view of the DFT-optimized geometry corresponding to the site experimentally observed in Figure 2d. (c) Simulated constant-height AFM image of this adsorption configuration showing the three isoindole units tilted slightly from the surface, while the fourth, almost invisible one, is closer at the surface.

To corroborate the experimental findings, we have performed dispersion-corrected DFT calculations. Figures 3a and 3b show the relaxed configuration (with the molecule initially in the configuration derived by ncAFM), with the Cu directly above an oxygen atom (Cu–O distance 260 pm) and the orientation of the molecule along the $\langle 1\bar{1}0 \rangle$ direction. Rotating the molecule away from this axis by ±5° results in an increase of the energy by approximately 150 meV, and an unconstrained optimization returns the molecule to its equilibrium orientation. Furthermore, the calculated geometry concurs with the interpretation of the experimental AFM results of Figures 2b and 2c, showing that only one of the four rings (pointing to the right in this orientation, Figure 3, indicated by the white curve) is tilted towards the substrate. Compared to a hypothetical flat CuPc geometry, the tilted configuration is stable by more than 1 eV (see Fig. S12 in the Supplementary Information). The simulated AFM image shown in Figure 3c (within the probe-particle model, see Methods Section) reproduces the characteristic appearance of the tilted molecule seen in the experiment and also clearly shows that the downward-bent ring appears dark, while the other three rings appear as bright protrusions.

**Pairs and chains of molecules**

In the experiments, CuPc pairs are observed already at low coverages. Here, both molecules occupy the adsorption site described above in neighboring substrate unit cells and also have the same orientation with respect to the surface. Continuing the growth up to 0.4 ML results in the formation of short molecular chains evolving from the pairs (see 'chains' in Figure 1 and Fig. S8 in the Supplementary Information). Each orientation of the CuPc molecule can form chains along only two of the three $\langle 1\bar{1}0 \rangle$ directions without molecular overlap. However, these two differently oriented chains are not symmetric with respect to the substrate; between two molecules there is either a region of the surface featuring In(6c) or In(5c). The majority of the chains are found with In(5c) atoms between the molecules (Fig. S9b). At low CuPc coverages, the length of the chains is limited by the amount of deposited material while at high coverages, the differently oriented chains lack long-range order and hinder each other from growing in length.

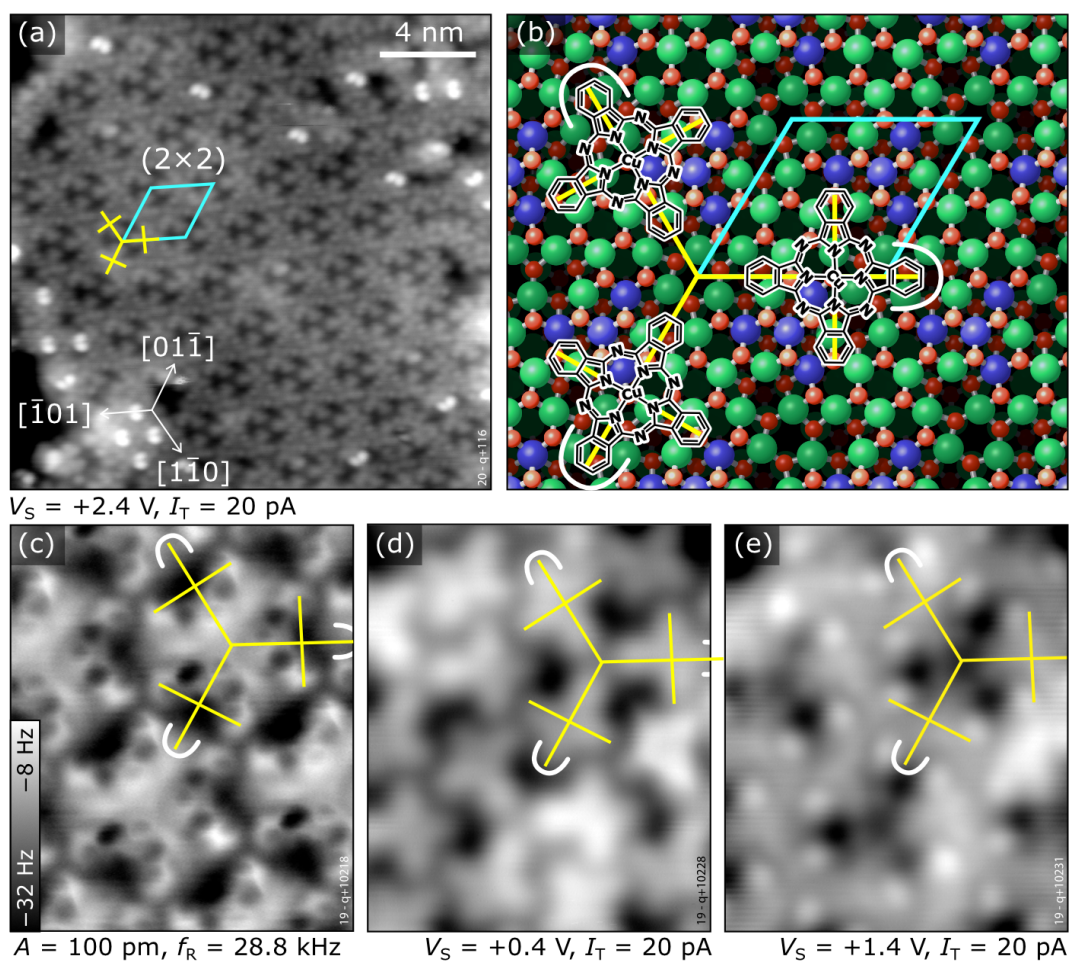

Fig. 4: The CuPc (2 × 2) superstructure on In$_2$O$_3$(111). (a) STM image showing a terrace covered by the (2 × 2) structure with the superstructure indicated. (b) Sketch of the (2×2) building block indicated with yellow crosses. The isoindole units bent towards the surface are indicated by white curves. (c) nc-AFM images reveal that the isoindole unit bent towards the surface points away from the center of the building block in each molecule (dark in AFM). (d, e) STM images of the same surface region as (c) acquired at different bias voltages, (d) in the HOMO-LUMO gap and (e) tunneling into the LUMO of the molecules. The images were acquired at (a) 80 K, and (c–e) 5 K.

## The (2 × 2) phase

Above a coverage of 0.4 ML, small patches of a new CuPc phase develop, completely covering the surface at a coverage of 0.75 ML. The arrangement shows a (2 × 2) superstructure with respect to the In$_2$O$_3$ unit cell, see Figure 4a, and the appearance in STM at some bias voltages is visually reminiscent of the kagome lattice, albeit with a lower symmetry ($C_{3v}$ instead of $C_{6v}$, see Fig. S10 in the Supplementary Information). The (2 × 2) unit cell contains three differently-oriented CuPc molecules that keep the adsorption site, orientation, and bent geometry of the isolated molecules. Figure 4b displays a sketch of the (2 × 2) structure, showing that the bent-down benzene rings are at the outer ends of the threefold star. In Figure 4, individual molecules are indicated by a cross-like symbol connecting opposite isoindole units; the long arms of these crosses connect the three molecules of the building block. Figure 4c shows an AFM image of the (2 × 2) structure. Imaging the same region of the surface with STM at a bias voltage within the HOMO-LUMO gap of the CuPc molecule, shown in Figure 4d, allows for an easier identification of the molecules. The benzene rings of the three molecules that are pointing at each other appear longer. Increasing the bias voltage to +1.4 V, *i.e.,* tunnelling into the LUMO of the CuPc molecules, is shown in Figure 4e. The CuPc molecules are imaged as featureless objects with bright protrusions located asymmetrically at the benzene rings (similar to the appearance of single molecules, Figure 1a) and the benzene rings pointing towards the center of the three-fold star appear darker. The bright double features (e.g., in the lower left part of Figure 4a) are excess molecules arranging in the structure described next.

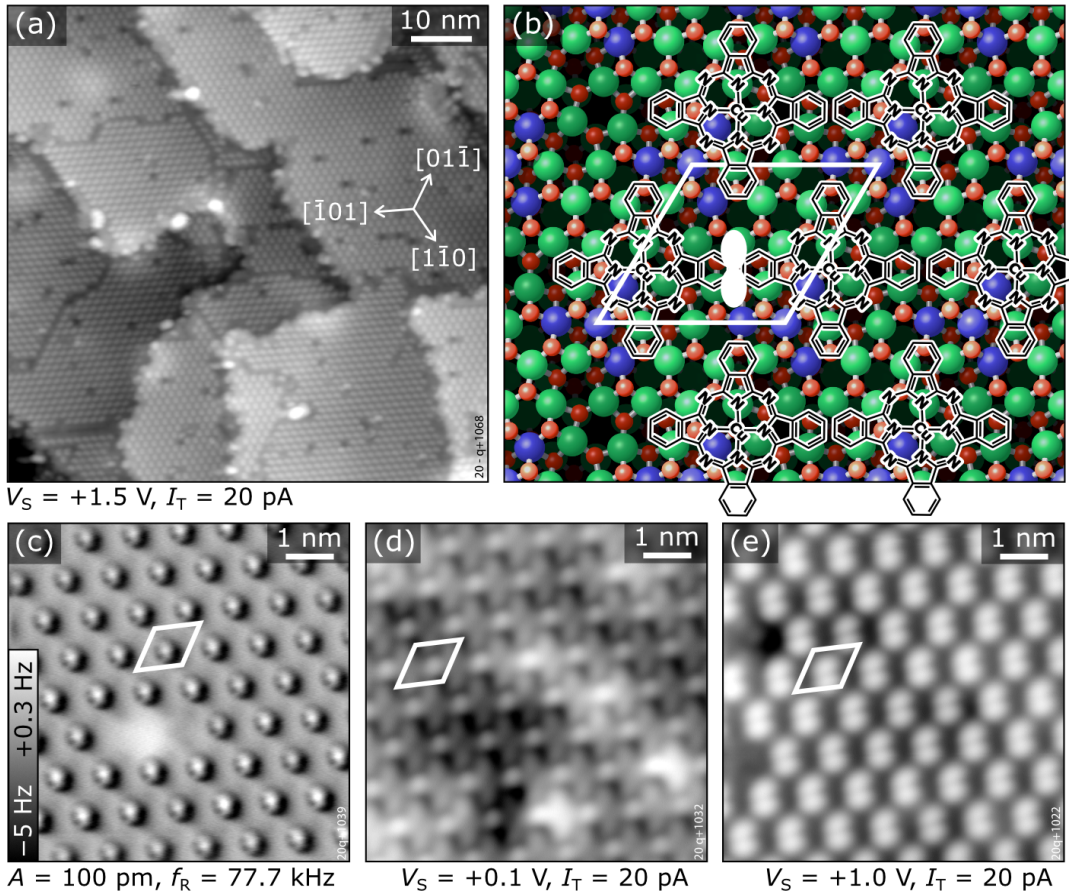

Fig. 5: The (1 × 1) superstructure of CuPc/In$_2$O$_3$(111). (a) Overview STM image showing the three distinct orientational domains separated by domain boundaries (dark lines). Dark dots within the domains are missing molecules. (b) Sketch of the In$_2$O$_3$(111) substrate superimposed with the structure of CuPc molecules in the (1 × 1) arrangement (one rotational domain is shown). The unit cell of the structure is indicated in white and the white double-lobed area indicates the overlap of neigboring molecules leading to maxima in the STM images. (c) nc-AFM image of an area with a single CuPc missing. The protrusions forming a (1 × 1) pattern are benzene rings bent away from the surface; the macrocycle of the molecule is not visible. (d, e) Empty-states STM images of the same rotational domain as in (c) but taken in an area without a CuPc vacancy. Panel (d) was acquired at a bias voltage below the CuPc LUMO; the cross-like features are individual CuPc molecules. The overlapping benzene rings of neighboring molecules are brighter than the rest of the molecule. Panel (e) displays the same surface area as (d) but at the energy of the LUMO. The bright double-lobed features originate from the LUMO at the benzene rings bent upwards, as also indicated in panel (b). All images were acquired at 5 K.

**Monolayer coverage**

Increasing the CuPc coverage to 1 ML (nominally one molecule per surface unit cell) leads to another change in the molecular arrangement, resulting in a CuPc structure with (1 × 1) symmetry present in three rotational domains (see Figure 5a). Domain boundaries are preferentially oriented along the $\langle 1\bar{1}0 \rangle$ directions. Constant-height AFM (Figure 5c) shows an array of hexagonally arranged features protruding from the surface; this indicates a bent geometry of the molecules different to what was observed for lower coverages. The macrocycle of the molecules is not visible any more as it is now further away from the tip in the constant-height images.

Imaging with an STM sample bias/energy within the CuPc HOMO-LUMO gap in an intermediate coverage regime, where the (1 × 1) and (2 × 2) structures coexist, allows us to determine the adsorption site and orientation of molecules (see Fig. S11 in the Supplementary Information). It is found that the adsorption site of the molecules remains the same as for lower coverages, *i.e.*, the Cu atom is above one O(γ) and also the azimuthal orientation remains the same, as shown in Figure 5b. The (1 × 1) structure contains uniformly oriented molecules in each domain, with one CuPc molecule per substrate unit cell *i.e.*, a density of $5.64 \times 10^{13}$ cm$^{-2}$. The uniform orientation and identical site of all CuPc molecules requires an overlap of the benzene rings of the molecules, however (Figure 5b). In STM, this overlap is best visible when imaging at the LUMO energy, shown in Figure 5e, where the nodal structure of the upwards bent orbital is imaged as bright double-lobed feature, also indicated in the sketch in Figure 5b. Tunneling into the HOMO-LUMO gap allows easy identification of the individual molecules, see Figure 5d.

To confirm the geometry, DFT calculations were performed on the (1 × 1) configuration derived from the experimental results. After relaxation, a structure with overlapping benzene rings is obtained, see Figure 6; the upwards bending due to the overlap is clearly visible in the side view (Figure 6b). In this (1 × 1) geometry, the adsorption energy is –2.71 eV per molecule, slightly weaker than for isolated molecules (–2.92 eV). The simulated constant-height AFM image shown in Figure 6c corroborates the experimental interpretation; only the upwards tilted benzene is visible.

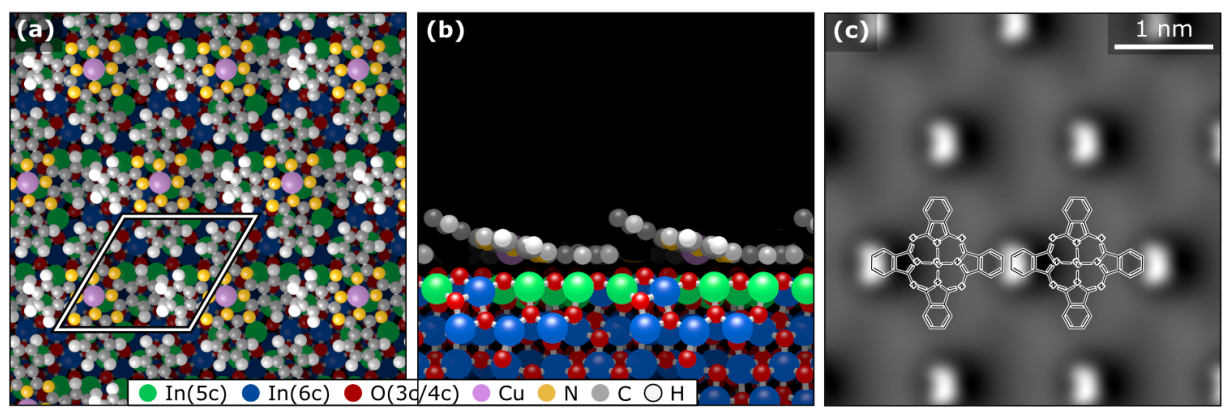

Fig. 6: Densely packed (1 × 1) layer of CuPc on $In_2O_3$(111). (a) Top view of the DFT-optimized geometry. (b) Side view of the DFT-optimized geometry. Note that only a single row of molecules is shown to clearly display the distortion of the molecules due to the steric hindrance between the benzene rings of neighboring molecules. (c) Simulated constant-height AFM image using the probe-particle model. Only the protruding parts of the molecules are visible.

**Discussion**

Investigating ordering of organic molecules on TCO surfaces in a bottom-up approach and under ideal conditions as presented in this work is important to understand and rationally tailor these systems for applications. While CuPc is found to grow upright standing due to dominant molecule-molecule interaction on rough ITO surfaces,[51,71] flat-lying adsorption is expected for atomically flat surfaces, maximizing the substrate-molecule interaction.[25,72,73] It also has been shown that oxygen deficiency by sputtering induces lying adsorption on ITO.[71] Both the ordering and the orientation of the molecules are key to device performance as they impact charge injection across the electrode-organic interface and charge transfer within the organic layer, in addition to the electronic level alignment. A well-defined ITO/organic (or $In_2O_3$/organic) interface and self-assembly of the molecules into an ordered first layer is helpful to propagate crystallinity into the organic film grown on top.

Phthalocyanine and $In_2O_3$(111) have different symmetries, 4-fold versus 3-fold, but the sizes of the molecule and the surface unit cell are similar. Thus, growing CuPc on $In_2O_3$(111) has the potential of resulting in a layer with one CuPc per surface unit cell, where the flat-lying molecules are anchored to the surface lattice, sufficiently spaced to avoid steric hindrance but also densely packed due to the lattice imposed from the substrate surface. This is, however, not what we have observed in this work. Instead, the adsorption geometry of individual molecule seems to be surprisingly robust and does not change when increasing the coverage from individual molecules scattered across the surface to the first monolayer, i.e., at a density where each surface unit cell is occupied by one CuPc. However, keeping the same geometry also in the full monolayer comes at the cost of partially distorting the molecules because the square shape of CuPc does not fit into a 3-fold (1 × 1) structure without overlap. In general, out-of-plane distortions of conjugated organic molecules are not uncommon and have been observed for phthalocyanines adsorbed on metal surfaces, e.g., by x-ray standing waves.[74-77]

In an earlier study[25] we have investigated the related molecule CoPc. It is instructive to consider these prior results, where everything is kept the same except the central metal ion being $Co^{2+}$ instead of by $Cu^{2+}$. For CoPc, we found two stable adsorption minima, which we denoted as "F" and "S" positions.[25] The "F" site is the adsorption site described here for CuPc. For a molecule in site "S" the central metal atom would be located above a neighboring In atom and the whole molecule is rotated clockwise by ~14°. These CoPc/In$_2$O$_3$(111) sites differed in adsorption energy obtained from DFT by 370 meV, with "F" being preferred. Still, for CoPc both sites were equally occupied, albeit with a slight preference for site "F". The experimental results on CuPc, described above, clearly show that CuPc is found only in site "F". The adsorption at the "F" site is found to be energetically more stable than the "S" geometry by 180 meV at the PBE level, which is mostly related to a larger van der Waals attraction in "F" geometry. Despite the energy difference being substantially smaller than for CoPc, where both adsorption sites are occupied,[25] in our experiments CuPc almost exclusively show adsorption on the (energetically more stable) "F" site already at low coverages. Also at higher coverages, monodisperse structures are observed for CuPc, in contrast to the mixture of structures formed by CoPc molecules in "F" and "S" sites. DFT calculations indicate that the energy difference between the "F" and "S" sites is robust and not related to the choice of the functional (the adsorption energy difference is 180 meV at the PBE level and 240 meV at the HSE level) or doping (up to a free charge carrier concentration of ca. $10^{18}$ e/cm³, the absolute adsorption energies change by less than 10 meV). We therefore consider it likely that the difference between CoPc and CuPc comes from different diffusion barriers: If CuPc (with a weaker metal–substrate bond) has a lower diffusion barrier than CoPc, CuPc is more likely to find the minimum-energy geometry "F".

The flat-lying adsorption geometry of CuPc on In$_2$O$_3$(111) with a homogeneous adsorption site linked to the substrate lattice is indicative of a strong interaction with the substrate. This is particularly important for charge injection across the oxide-organic interface. Moreover, the overlap of neighboring molecules along the ⟨1$\bar{1}$0⟩ directions in the (1 × 1) structure has the potential of an enhanced 2-dimensional charge transport within the CuPc layer, which is a topic that requires further investigation. The stacking observed in chains or in the (1 × 1) structure (but without overlap) is a typical arrangement of (M)Pc molecules on various surfaces, including metals,[78–80] oxides,[48] and graphene,[81,82] as it allows dense packing of the cross-shaped molecules. Note that on metals such an arrangement is often a coincidence structure, while on oxides, adsorption sites are usually more strongly linked to the surface lattice.

Studies of well-ordered first layers of organic molecules on oxide surfaces are scarce. Another promising system for a well-defined first monolayer is CoPc/CoO(111)[48] where a densely-packed structure with two adsorption sites within the superstructure was found; a drawback is the domain size of approximately only 10 nm. On rutile $TiO_2$, the material studied most of all oxide surfaces, MPcs (and also Zn-tetraphenyl porphyrin TPP) struggle to form well-ordered structures in the first layer, but the layer on-top of either MPc or TPP is surprisingly well-ordered with two different structures coexisting for lying molecules that change into upright-standing molecules after annealing at higher temperature[30,33,34] In contrast to the MPc growth on $TiO_2$, we observe a well ordered first layer with domains that are larger than those formed by CoPc on CoO(111). This densely packed first monolayer formed by CuPc on $In_2O_3$(111) thus serves as a promising foundation for rationally designing crystalline CuPc thin films.

**Conclusion**

This study outlines the structural evolution of CuPc on $In_2O_3$(111) as coverage increases. At low coverages, individual molecules and pairs of molecules are observed; with increasing coverage they grow into short chains. Eventually, the formation of large and uniform CuPc domains occurs with (2 × 2) and (1 × 1) structures at 0.75 and 1 ML, respectively (1 ML ≡ 1 CuPc per substrate unit cell). The latter shows long-range order in spite of significant overlap between neighboring molecules. Across all coverages from isolated molecules to the full (1 × 1) monolayer, the CuPc is adsorbed in the same three symmetry-equivalent configurations. This substrate-driven ordering prompts a conformational change in the isoindole units — from relatively flat structures at lower coverages to strong upwards bending of one of the benzene rings due to the overlap at monolayer coverage — demonstrating the adaptive nature of molecular layers in response to the interplay between substrate-molecule and intermolecular interactions.

**Conflicts of interest**

There are no conflicts to declare

**Data availability**

All relevant data are within the manuscript and Supporting Information. Additional supporting data are available from the corresponding author upon request.

**Acknowledgement**

This research was funded or in part by the Austrian Science Fund (FWF) [10.55776/V773, 10.55776/Y1157, 10.55776/COE5]. For the purpose of open access, the authors have applied a CC BY public copyright license to any Author Accepted Manuscript version arising from this submission. MB acknowledges financial support through the ERC and MEYS CR co-founded IMPROVE V project CZ.02.01.01/00/22_010/0002552; UD through the ERC-2019-ADG, Project #883395 "WatFun"; OTH through the FWF-START-project Y1157 "MAP-DESIGN"; M.E. through the Marie Skłodowska-Curie Actions (Project 101103731, SCI-PHI).